\documentclass[manuscript]{aastex} 

\lefthead{Tokunaga et al.}
\righthead{Multiplicity of AS 353  }

\usepackage{graphicx}
\usepackage{amssymb}
\begin{document}


\title{A Subarcsecond Companion to the T Tauri Star AS 353B\altaffilmark{1}}

\author{
A. T. Tokunaga,\altaffilmark{2} 
Bo Reipurth,\altaffilmark{2}
W. Gaessler,\altaffilmark{3}
Yutaka Hayano,\altaffilmark{4}
Masahiko Hayashi,\altaffilmark{5}
Masanori Iye,\altaffilmark{4}
Tomio Kanzawa,\altaffilmark{5}
Naoto Kobayashi,\altaffilmark{6}
Yukiko Kamata,\altaffilmark{4}
Yosuke Minowa,\altaffilmark{6}
Ko Nedachi,\altaffilmark{5}
Shin Oya,\altaffilmark{5}
Tae-Soo Pyo,\altaffilmark{5}
D. Saint-Jacques,\altaffilmark{7}
Hiroshi Terada,\altaffilmark{5} 
Hideki Takami,\altaffilmark{5} 
Naruhisa Takato\altaffilmark{5}  }

\altaffiltext{1} {Based on data collected at Subaru Telescope, which 
   is operated by the National Astronomical Observatory of Japan.}

\altaffiltext{2}{Institute for Astronomy, University of Hawaii, 2680 Woodlawn 
   Drive, Honolulu, HI 96822;
   tokunaga@ifa.hawaii.edu, reipurth@ifa.hawaii.edu.}

\altaffiltext{3}{Max-Planck-Institut f\"ur Astronomie, K\"onigstuhl 17,
Heidelberg D-69117, Germany; gaessler@mpia-hd.mpg.de}

\altaffiltext{4}{National Astronomical Observatory of Japan, Mitaka, 
   Tokyo 181-8588, Japan; y.hayano@nao.ac.jp, iye@optik.mtk.nao.ac.jp,
   kamata@merope.mtk.nao.ac.jp}

\altaffiltext{5}{Subaru Telescope, 650 North A'ohoku Place, Hilo, HI  96720;
Firstname.Lastname@SubaruTelescope.org.}

\altaffiltext{6}{Institute of Astronomy, Graduate School of Science, University
of Tokyo, Mitaka-shi, Tokyo 181-0015, Japan; naoto@ioa.s.u-tokyo.ac.jp.}

\altaffiltext{7}{Groupe d'astrophysique, Universit{\'e} de Montr{\'e}al,
2900 Boul. {\'E}douard-Montpetit, Montr{\'e}al, Qu{\'e}bec, H3T 1J4 Canada;
dastj14@agora.ulaval.ca.}

\begin{abstract}

Adaptive optics imaging of the bright visual T Tauri binary AS~353
with the Subaru Telescope shows that it is a hierarchical triple
system.  The secondary component, located 5\farcs6 south of AS~353A, 
is resolved into a subarcsecond binary, AS~353Ba and Bb, 
separated by 0\farcs24.  Resolved spectroscopy of the two close components 
shows that both have nearly identical spectral types of about M1.5.  Whereas
AS~353A and Ba show clear evidence for an infrared excess, AS~353Bb does
not. We discuss the possible role of multiplicity in launching the large
Herbig-Haro flow associated with AS~353A.

\end{abstract}  

\keywords{stars: formation -- binaries: general -- ISM: jets and outflows}

\section{INTRODUCTION}

The T Tauri star AS~353A (HBC 292) was first recognized as an
H$\alpha$ emission object by Merrill \& Burwell (1950) and Iriarte \&
Chavira (1956). 
AS~353A is particularly interesting, partly because
it is one of the visually brightest known T Tauri stars ($V$ $\sim$ 12.7), 
but especially because it drives the prominent Herbig-Haro
(HH) flow HH~32 (Herbig 1974; Herbig \& Jones 1983; Mundt, Stocke, \& Stockman
1983; Solf, B\"ohm \& Raga 1986; Hartigan, Mundt, \& Stocke 1986;
Curiel et al. 1997). The optical spectrum of AS~353A is discussed in
detail by Herbig \& Jones (1983), B\"ohm \& Raga (1987), and
Eisl\"offel, Solf, \& B\"ohm (1990), who find that it is heavily
veiled with a rich emission-line spectrum and powerful H$\alpha$ 
emission. Furthermore,
AS~353A is detected in the radio continuum (Anglada et al. 1998), as
is common for HH driving sources. The star displays considerable
variability (e.g., Fernandez \& Eiroa 1996), and is also known as
V1352~Aql.

AS~353A has a fainter ($V$ $\sim$ 14.6) companion star, AS~353B (HBC
685), about 5\farcs6 to the south. In marked contrast with the many
studies of the A-component, this companion has been only poorly studied. It
is a weak-line T Tauri star with a spectral type estimated to be M0 
by Cohen \& Kuhi (1979) and M3 by Prato, Greene, \& Simon (2003).
Both stars are located in a cloud cavity,
whose edges are illuminated by the two stars, as seen well in the {\em 
Hubble Space Telescope} images of the region by Curiel et al. (1997). 

In this paper, we present near-infrared adaptive optics observations
of the AS~353A/B pair.  Our motivation for undertaking these observations
was to search for a companion to AS~353A.  We were unsuccessful in
this, but we found AS~353B to be a subarcsecond binary (independently 
discovered by White et al. 2002).

\section{OBSERVATIONS}

Diffraction-limited imaging of AS~353 A and B was obtained with the
Infrared Camera and Spectrograph (IRCS) instrument at the Subaru
Telescope (Tokunaga et al.  1998; Kobayashi et al.  2000) on 2001 
July 12 UT. The IRCS was used with the Subaru Telescope adaptive optics
system (Takami et al.  1998; Gaessler et al.  2002) using AS~353A as
the reference star for the wavefront sensor.  Images were taken at 
$H$ (1.63 \micron), $K$ (2.19 \micron), and $L'$ (3.72 \micron)  with the 
isophotal wavelengths are given in parentheses. The pixel scale was
22.560 $\pm$ 0.059  milliarcsec pixel$^{-1}$.  The Mauna Kea Observatories
near-infrared filters described by Simons \& Tokunaga (2002) and 
Tokunaga, Simons, \& Vacca (2002) were used.  The observations were 
made with a box-shaped five-image
dither pattern with a separation of 2\farcs0.  The total integration
time for the $H$, $K$, and $L'$ images were 100, 75, and 180 s,
respectively.

Although our objective was to determine if AS~353A is a binary star, 
we instead discovered that AS~353B is a binary.
There is no evidence in our images that AS~353A has a companion with a
separation greater than 0\farcs1.  Figure 1 shows the $L'$ image of
the AS~353 system.

The Strehl ratio was estimated from the ratio of the peak flux to that of the
total flux and comparing these values to a theoretical diffraction-limited
image.  The Strehl ratio was found to be 0.09 at $H$, 0.21 at $K$, and 0.51
at $L'$.

The images were reduced using IDL procedures. 
Aperture photometry was performed on AS~353A with a radius of
1\farcs6.  The conversion from ADU s$^{-1}$ to magnitudes was determined
from the zero point magnitude derived previously  
since a photometric standard was not observed due to time
limitations.  The airmass during the observations ranged from 1.16 to 1.28.
The sky conditions were good, and the atmospheric extinction 
coefficients from Krisciunas et al. (1987) were adopted.  
Relative photometry of the components of A, Ba, and Bb was
obtained using an aperture radius of 5 pixels (0\farcs11).  The
magnitudes for AS~353Ba and Bb were then obtained from that of AS~353A,
and the photometric results are shown in Table 1.

The uncertainties in Table 1 were obtained from the variations in the 
photometric counts from frame to frame.  Additional systematic uncertainties
arise from the use of the photometric standard of another night and the
adopted zero magnitudes based on the available photometric observations.
The systematic uncertainties are approximately 0.05 mag.

We obtained spectroscopic observations of AS~353Ba and Bb 
in the $H$ and $K$ bands on 2002 May 26 UT using the IRCS in the grism 
mode with a 0\farcs15 slit. The resolving power was 610 and 640 
at $H$ and $K$, respectively.    As with the imaging, the Subaru adaptive 
optics system was used, and  the
slit was aligned along the Ba and Bb components.  The object was nodded
along the slit by 3\farcs0 in two positions.  A set of four spectra
(two in the positive beam and two in the negative beam) was obtained
at both the $H$ and $K$ bands.  The total integration time was 120 s at
$H$ and 240 s at $K$.  The spectra were reduced using IDL routines and
the ``optspec'' procedure provided by M. 
Buie.\footnote{http://www.lowell.edu/users/buie/idl/idl.html}

The A0 star HD 189411 was used to correct for telluric absorption.  
The typical method of obtaining the spectrum of an object 
is to divide by the telluric standard and multiply by a 
Planck function that approximates the continuum of the standard. 
However, we employed an
alternative method described by Vacca, Cushing, \& Rayner (2003) of using the known 
spectral type of the telluric standard and the measurement of at least one of its 
absorption lines to model both the continuum and line absorption.  The 
instrumental profile and atmospheric absorption can then be derived, and 
these are used to obtain the spectrum of AS~353Ba and Bb.   The
primary advantage of this method is that the spectral lines in the telluric standard
are removed to a greater degree than is possible by other methods.
The resulting spectra are shown in Figure~2.

\section{DISCUSSION}

\subsection{The AS353B Binary}

Figure 1 shows the $L'$ image of AS~353A/B, and it is apparent that
AS~353B is resolved into a subarcsecond binary, where we denote the
brighter component as Ba and the fainter as Bb. It has a separation of
0\farcs24 $\pm$ 0\farcs01 with a position angle of 107\fdg2 $\pm$
1\fdg3 as measured from AS~353Ba. The separation of AS~353A and Ba is
5\farcs63 $\pm$ 0\farcs01 with a position angle of 174\fdg1 $\pm$
0\fdg2 as measured from AS~353A.  The uncertainties in separation 
are derived from the standard deviation of the individual frames.  
The uncertainty in the position angle does not include  
the uncertainty in mounting the instrument onto the
telescope (estimated to be $\pm$ 0\fdg1).  These values improve 
upon those of Chelli, Cruz-Gonzalez, \& Reipurth (1995), since the A--B
separation is more precisely measured now that we have resolved the
components of AS~353Ba and Bb.

The spectral type of AS~353A is K2 (Basri \& Batalha 1990).
Prato et al. (2003) show a spectrum
of AS 353A in which He I and strong Br$\gamma$ emission is seen.
The spectrum of AS~353Ba and Bb 
combined is  M0--M3 (Cohen \& Kuhi 1979; Prato et al. 2003), and as can 
be seen in Figure 2,  the spectra of AS353~Ba
and Bb are very similar.  We compare the spectrum of AS353~Ba to 
the standard star spectra of  Wallace \& Hinkle (1997) and Meyer 
et al. (1998) in Figures 3 and 4, respectively.  For this comparison, 
we have converted to F$_\lambda$ (W m$^{-2}$ $\mu$m$^{-1}$)
and wavelength units, and smoothed the spectra to match that of our
AS~353Ba spectrum.  

Figure 3 shows that in the $K$ band the spectrum of AS~353Ba resembles that
of an M dwarf star since the CO absorption at 2.3--2.4 $\mu$m is too
strong in the spectra of the giants.  The Mg I line at 2.28 $\mu$m is a 
good indicator of the spectral type.  We see that this line is nearly absent 
by spectral type M2V.  $K$-band spectra obtained by J. Rayner 
(2003, private communication) show that
AS~353Ba is similar to a M1.5V main-sequence star but not later.  This
conclusion is also consistent with comparison to the spectral standards
shown by Ali et al. (1995).  Our spectral type estimate is a little earlier but
consistent with the spectral type of M3 $\pm$ 2 obtained by Prato et al. (2003).

Figure 4 shows that the spectrum of AS~353Ba is not a good match
to the spectra of either dwarf or giant stars.  The Mg I (1.58 $\mu$m) and CO
band (1.62 $\mu$m) are best matched to a giant spectrum earlier than M1.
However the Mg I (1.71 $\mu$m) line is best matched to a dwarf 
spectrum of M1.5V or later.  Thus the $H$-band spectrum shows spectral 
characteristics of both dwarfs and giants.  Evidence that premain-sequence
stars show spectral characteristics intermediate between dwarfs and giants
has also been presented by Luhman (1999).

The straight continuum at 2.0--2.5 \micron\ is also an
indication of an early M spectral type.  Comparison of the continuum
shapes to the M dwarf spectral library of Leggett et al. (2002)\footnote{ 
Spectra available at anonymous ftp site: ftp.jach.hawaii.edu, 
cd /pub/ukirt/skl.  }
indicates that the spectral type is about M1 or later.  Therefore, our
best estimate is that the spectral types of both AS~353Ba and Bb are
about M1.5.

The $H-K$ and $K-L'$ colors can be used to roughly classify the
objects in terms of infrared excess, extinction, and spectral type.
In Figure 5 we show the position of AS~353A, Ba, and Bb in a color-color
diagram.  We see that AS~353A and Ba have infrared excesses 
and follow the classical T Tauri line derived by
Meyer, Calvet, \& Hillenbrand (1997).  The difference in brightness
between Ba and Bb may result from the infrared excess of component~Ba, or
alternatively from variability.

Our magnitudes for AS~353A at $H$ and $K$ are within $\pm$0.3 
of those presented by Cohen \& Schwartz (1983) and by Prato et al. (2003), 
and our $L'$ magnitude  is 0.1 brighter than found by Cohen \& Schwartz (1983).
We note that Fernandez \& Eiroa (1996) show that the visual magnitude of
the AS~353 system (A, Ba, and Bb) varies by $\pm$0.5 mag.
The combined magnitude of Ba and Bb are within $\le$0.07 mag of
that reported by Cohen \& Schwartz (1983) and within 0.3 mag of
Prato et al. (2003).    The $H-K$ color of AS 353A and  AS 353B (combined)
are within 0.05 and 0.1 mag, respectively, of the colors derived
from the data of Cohen \& Schwartz (1983) and by Prato et al. (2003).

AS~353Bb has colors similar to that of an M0 star.  
Combined with the spectral type of AS~353Bb
obtained above, we infer that the extinction of AS~353Bb is nearly zero
in the near infrared.  Note that Prato et al. (2003) obtained a visual 
extinction of 2.1 $\pm$ 0.8 for the combined light from Ba and Bb.
However, if we apply such a correction to AS~353Bb, the resulting colors 
would be inconsistent with the early M spectral type found above.  Thus
the visual extinction value obtained most likely pertains to Ba, the
brighter component.
   
Cohen \& Kuhi (1979) reported that the H$\alpha$ equivalent widths for 
AS 353A and AS 353B (combined) were 124 \AA\ and 4.4 \AA\, respectively.
AS 353A is redder and has more thermal emission than the binary companion, 
and within the binary system, AS 353Ba has detectable 
thermal emission but AS 353Bb does not.  Coupled with the H$\alpha$
emission, this is evidence for emission from an accretion disk in 
both AS 353A and AS 353Ba.  
 
AS 353A is clearly the most active member of the system, since the 
outflow is emanating from this component and it has strong emission
lines.  Duch\^{e}ne et al. (1999) found evidence that, in binary systems where
both stars are still accreting material, the more massive star has
the larger accretion rate as measured by the H$\alpha$ luminosity.
This is consistent with the picture that AS 353A is the most
massive object in the system and that it has a very red $K-L'$ 
color due to an accretion disk. The AS 353 system fits
the trend reported by Duch\^{e}ne et al. (1999) 
that the more massive object has the higher accretion rate in binaries.
 
The distance to AS353 is uncertain.  Herbig \& Jones (1983) estimated 
the distance to be 300 pc, but they emphasized the uncertainty of this 
determination and noted that their astrometric distance methods tend to 
overestimate the derived distances.  Prato et al. (2003) adopt a distance
of 150 $\pm$ 50 pc following the estimated distances of Edwards \& Snell
(1982) and Dame \& Thaddeus (1985), who obtained distances of
$\ge$150 pc and 200 $\pm$ 100 pc, respectively.  However, the Edwards \& 
Snell distance estimate does not take into account the now-known 
binarity of AS~353B, and they also assume that the AS~353B component
is a single main-sequence star.  The Dame \& Thaddeus distance estimate is 
for that of the Aquila Rift.  They note that the distance estimates range from 
150 pc near $\ell$ = 20$^\circ$ to $\sim$300 pc near $\ell$ = 40$^\circ$.
Since AS~353A is at $\ell$  = 46.5$^\circ$ and the visual extinction is
about 2.1--2.9 (Prato et al. 2003; Cohen \& Kuhi 1979), it is likely that
AS~353A is in the foreground to the Aquila Rift and at a distance of 
$\le$ 300 pc.  Prato et al. (2003) find that a distance of 150 pc gives a
reasonable age for the AS353B component, and this is the best
estimate of the distance at this time.

For a distance of 150 pc, the absolute $K$ magnitude of AS~353Ba and Bb
separately is approximately 3.2 mag.  This is about 2.6 mag brighter than the
absolute magnitude of a main-sequence M2V star.  

The projected separation between AS~353Ba and Bb is only 36 AU.
It is interesting to note that Jensen, Mathieu, \& Fuller (1996) 
find that binaries with
separations between a few and about 100 AU have low submillimeter continuum
emission.  This suggests to them that such binaries do not have circumbinary
disks.  The large $K-L'$ color for AS~353A and small $K-L'$ for AS~353B
is consistent with the results of Jensen et al. (1996).  In addition, 
White et al. (2002) find that close binaries in triple systems are always the 
nonaccreting component.  In the case of AS 353, there is some
accretion in AS 353Ba, but it is small in comparison to AS 353A.

\subsection{An Evolutionary Scenario}

AS 353A is associated with the well-known HH 32 flow\footnote{
The HH 32 flow has an inclination of about 70$^\circ$ with respect to
the plane of the sky, and so is significantly foreshortened
(e.g., Herbig \& Jones 1983; Curiel et al. 1997). 
A faint group of HH knots, HH~332, is found 68$''$ SSW of AS~353A
(Davis, Eisl\"offel, \& Smith 1996), and it could well be the fading
remnants of a terminal working surface of the HH~32 flow.  However,
there is an approximately 40$^\circ$ angle between lines from AS~353A
to HH~32 and HH~332, which is a substantial difference, even
considering that HH flow axes are known to vary with time.  But when
we take into account the 70$^\circ$ inclination of the HH~32 flow, the
angle between the actual directions from AS~353A to HH~32 and HH~332
reduces to about 13$^\circ$ degrees, which is commonly seen in giant
HH flows.  If this is so, the physical length of the HH~32/HH~332
large-scale flow, assuming a distance of 150~pc, is 30,000~AU, or
0.15~pc.  While this would in itself not be considered a giant flow,
it does indicate that the HH~32 flow may be much larger than its
currently recognized extent would suggest.  Since HH~32 is a
redshifted HH object, a prediction of the connection with HH~332
would be that this HH flow should also be redshifted. In contrast, the large
majority of HH flows are blueshifted, due to selection effects related
to dust obscuration.
}.
In the last few years, it has been recognized that HH flows may
commonly attain very large, parsec-scale dimensions (e.g., Reipurth,
Bally, \& Devine 1997; G\'omez, Kenyon, \& Whitney 1997). Such giant
HH flows provide a fossil record of the mass-loss activity and
accretion history of their driving sources. Detailed studies of 14
sources of giant HH flows have revealed an {\em observed} binary
frequency of about 80\%, of which half are higher order multiples,
leading Reipurth~(2000) to postulate the stellar dynamics jet
hypothesis, in which dynamical decay of triple or multiple systems
leads to strong outflow activity that manifests itself in the
multiple shock structures observed in giant HH flows.  

The stellar dynamics jet hypothesis 
predicts that the HH~32 source should be part of a triple or multiple
system that has recently gone from a nonhierarchical configuration
to a hierarchical one.  Our observations indeed show AS~353A/B to be a
hierarchical triple system.  However, since the HH~32 flow
unquestionably arises in AS~353A, and not in AS~353B,  
we would expect AS~353A to be a binary as well.

We thus speculate that the AS 353 AB system is not only a triple
system, as observed, but also a quadruple, with AS~353A being
an unresolved close binary. If so, this would make it very similar to
UZ Tau, where one component is a close visual binary (Simon et
al. 1992) and the other is a spectroscopic binary (Prato et
al. 2002). This configuration can arise from an initial
nonhierarchical quadruple system. Such systems are unstable, and
subsequent dynamical interactions would transform this system within
about a hundred crossing times (Anosova 1986; Sterzik \& Durisen
1995). This can occur in one of several ways.  Ejections can lead to
(1) the escape of two stars; (2) one star escaping and another placed
in a distant orbit; (3) a binary being placed in a distant orbit. In all
such scenarios, the most massive member remains a
binary that becomes tighter by the ejection process. The members that
either escape or are placed into a distant orbit will likely have some
of their disks truncated in the process, or at least will be displaced
from the center of the potential well, leading to less accretion. In
either case the outlying component(s) should have less circumstellar
material than the inner dominant binary (e.g., Bate \& Bonnell 1997;
Bonnell et al. 2001; Bate, Bonnell, \& Bromm 2002).

It is tempting to interpret AS~353, and other premain-sequence binaries like
UZ~Tau, in terms of case~3. If this is so, then  
AS~353A itself should be a close binary surrounded by substantial
amounts of circumstellar material. Figure 5 shows that AS~353A has a
very substantial infrared excess, in contrast to the Ba and Bb
components. Binary motion in such a viscous environment may lead to
rapid spiraling-in of the components, suggesting that at present the
binary would be much closer than the Ba/Bb pair, consistent with our
upper limit of 0\farcs1 (15 AU in projection) for the presence of any
companion.

\section{CONCLUSIONS}

\begin{enumerate}
	
\item Adaptive optics imaging of AS~353A reveals that its companion
5\farcs63 to the south is a close binary system with a separation of
0\farcs24.  AS~353A itself did not show any companion with a projected
separation larger than 0\farcs1.  Thus the AS~353 system is a 
hierarchical triple system.

\item The components of the close binary system, AS~353Ba and Bb, 
have nearly identical 
spectral types of approximately M1.5.  At an assumed distance of 150 pc, 
the absolute $K$ magnitude of Ba and Bb places them well above the
main sequence.  The spectra of Ba and Bb have spectral characteristics
of both dwarfs and giants.

\item AS~353A and Ba show infrared excesses that are typical for T
Tauri stars, but AS~353Bb does not.  The infrared colors and spectra of
AS~353Bb show that the extinction to this source is nearly zero.

\item We suggest that the present AS~353 system evolved dynamically
from an unstable nonhierarchical quadruple system and predict that
AS~353A is a close binary.

\end{enumerate}

\acknowledgments{We thank the staff and crew of the Subaru Telescope
for their invaluable assistance in obtaining the observations reported
here.  We are grateful to J. Rayner and S. Leggett, who made available 
their M dwarf  spectral libraries to us, and to W. Vacca and M. Cushing for 
providing their version of the `xtellcor' program prior to publication. 
We thank G. Herbig for helpful comments on the spectrum of AS 353A.
ATT was supported by NASA Cooperative Agreement no. NCC 5-538.  
}

\clearpage

\begin{table}[h]
\caption{Magnitudes for the AS353 System$^{a}$} 
\bigskip
\begin{tabular}{lrrr} \hline
Object & $K$ & $H-K$ & $K-L'$  \\
\hline 
AS353A  & 7.97 & 0.93 & 1.73 \\
AS353Ba  & 9.26 & 0.35 & 0.56 \\
AS353Bb  & 9.45 & 0.26 & 0.13 \\
\hline
\end{tabular}
\end{table}

\noindent (a) The 1 $\sigma$ uncertainties for $H$, $K$, and $L'$ are \\
0.03, 0.03, and 0.06 mags, respectively.


\clearpage


\begin{figure}
\epsscale{0.90}
\plotone{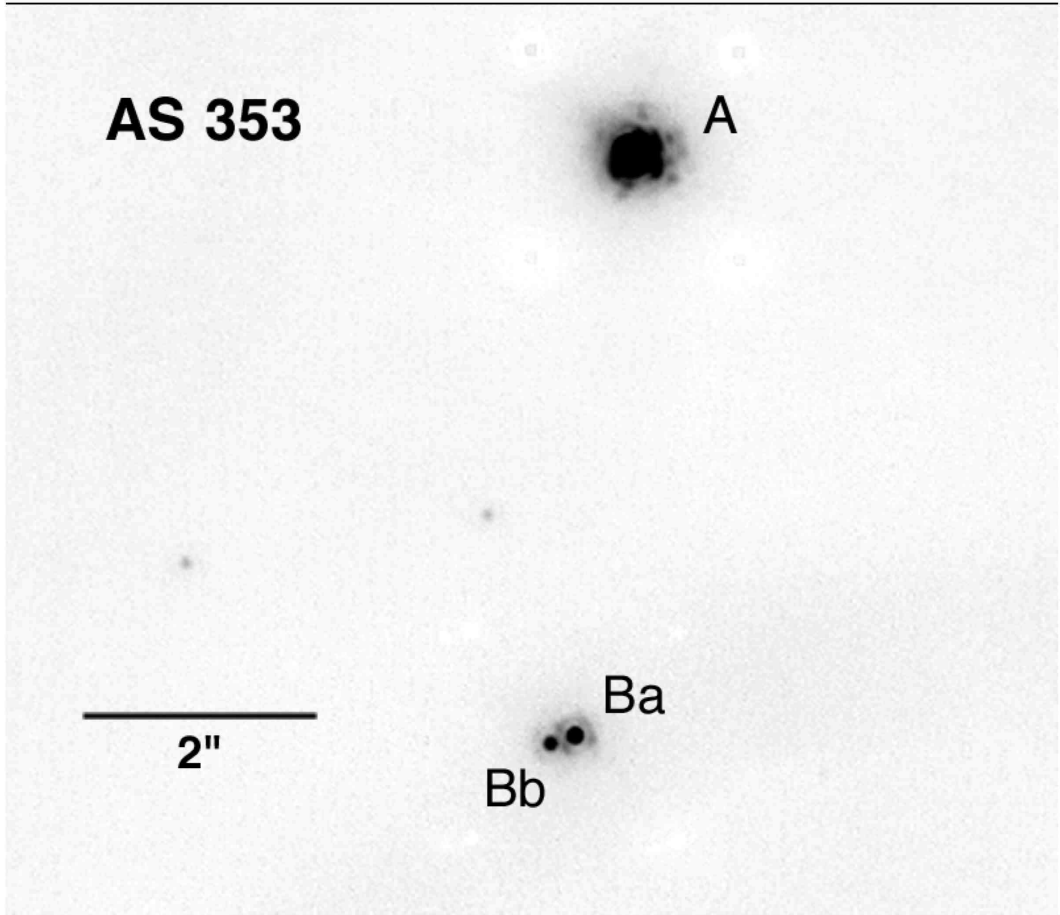}
\caption{$L'$ image of AS~353A, Ba, and Bb obtained at the Subaru Telescope 
with IRCS and the Subaru adaptive optics system.  North is up and East to the 
left.  Structure in the wings of the point-spread function of AS~353A is
speckle noise.  Since the integration time was short, no extended structure in 
AS~353A was observed.}
\end{figure}

\clearpage

\begin{figure}
\epsscale{0.85}
\plotone{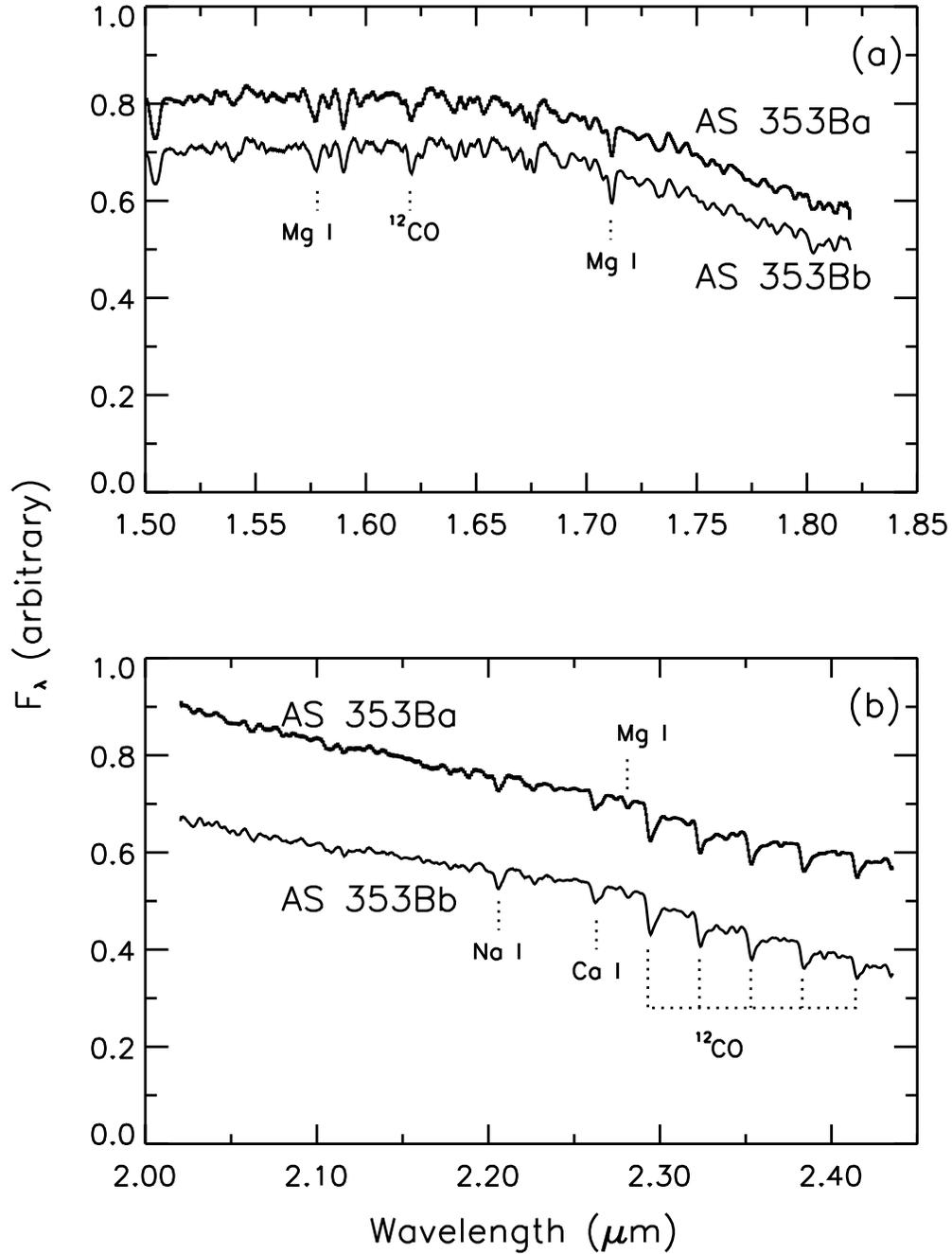}
\caption{Grism spectra of AS~353Ba and Bb. 
(a) The $H$-band spectra of AS~353Ba (upper curve) and AS~353Bb (lower curve). 
(b) Same as above for the $K$ band.}
\end{figure}

\clearpage

\begin{figure}
\epsscale{0.85}
\plotone{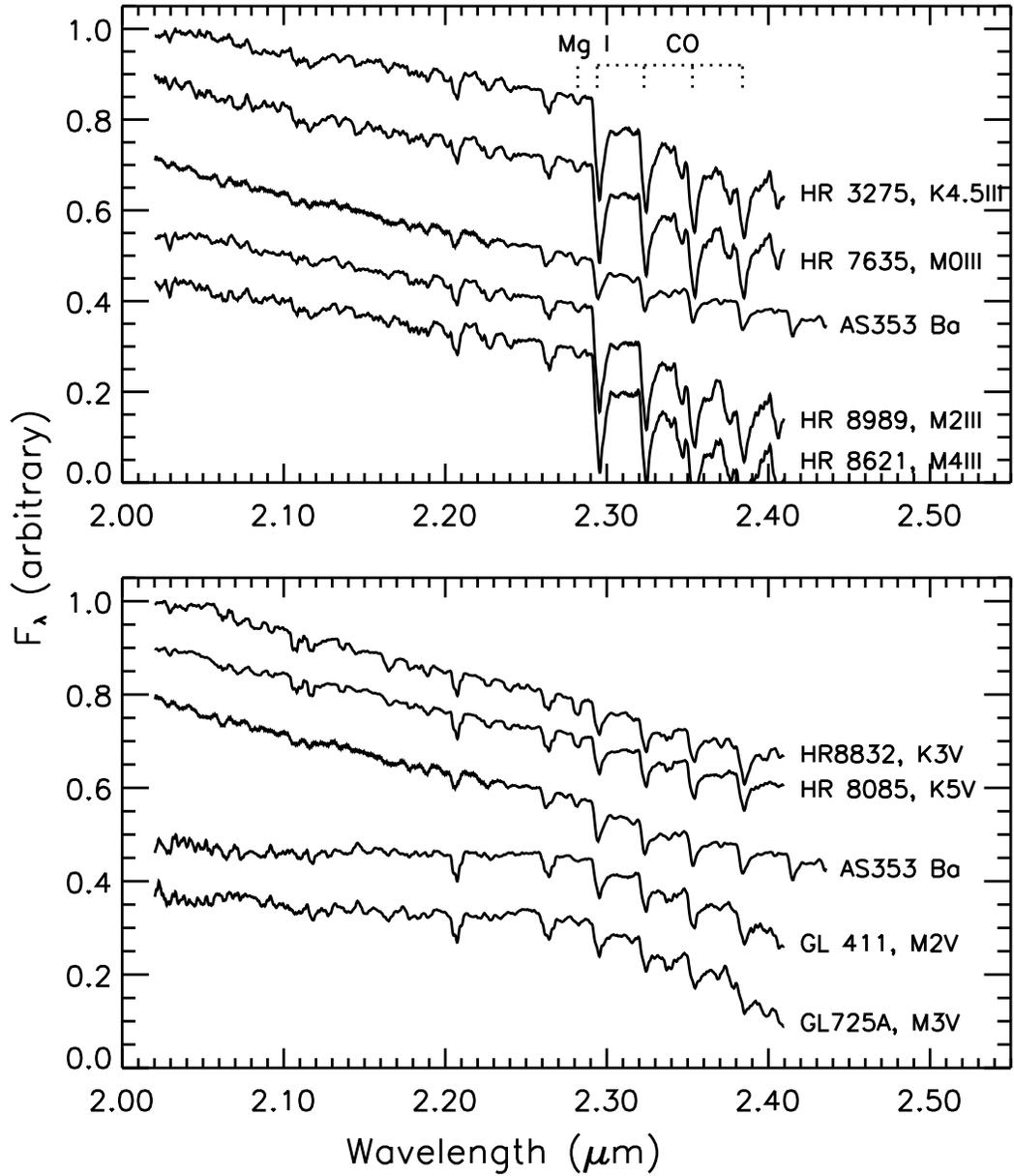}
\caption{Comparison of AS~353Ba with spectral standards from Wallace \& 
Hinkle (1997).  The comparison to giant main-sequence stars is shown in the 
top panel, and to dwarf stars in the bottom panel.
The Mg I line at 2.28 $\mu$m and the $^{12}$CO bandheads are shown.}  
\end{figure}

\clearpage

\begin{figure}
\epsscale{0.85}
\plotone{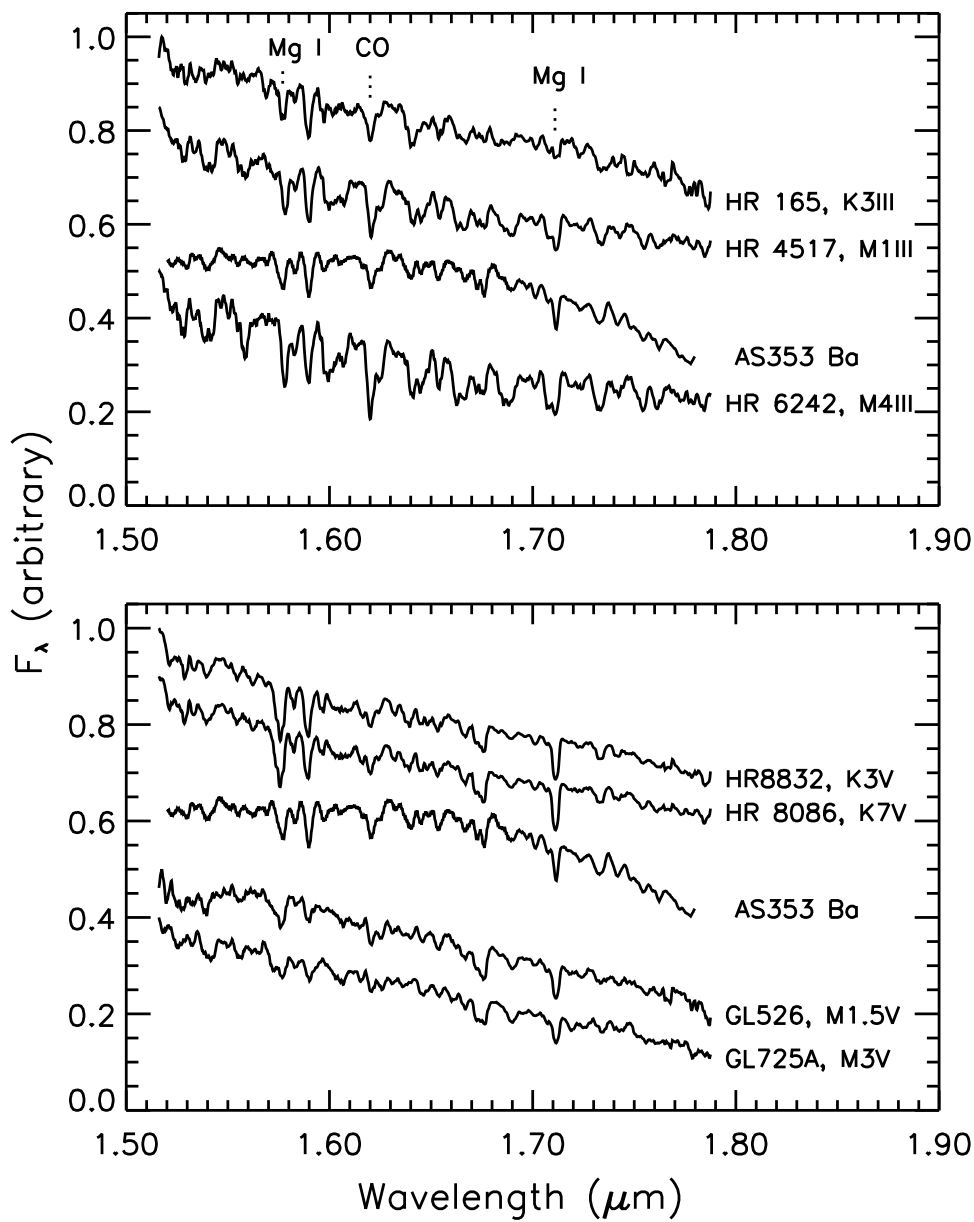}
\caption{Comparison of AS~353Ba with spectral standards from Meyer et al.
(1998).  The comparison to giant main-sequence stars is shown in the 
top panel, and to dwarf stars in the bottom panel.
The Mg I lines at 1.58 and 1.71 $\mu$m and the $^{12}$CO band at 1.62
$\mu$m  are shown.  The continua in the Meyer et al. spectra were
apparently flattened.}
\end{figure}

\begin{figure}
\plotone{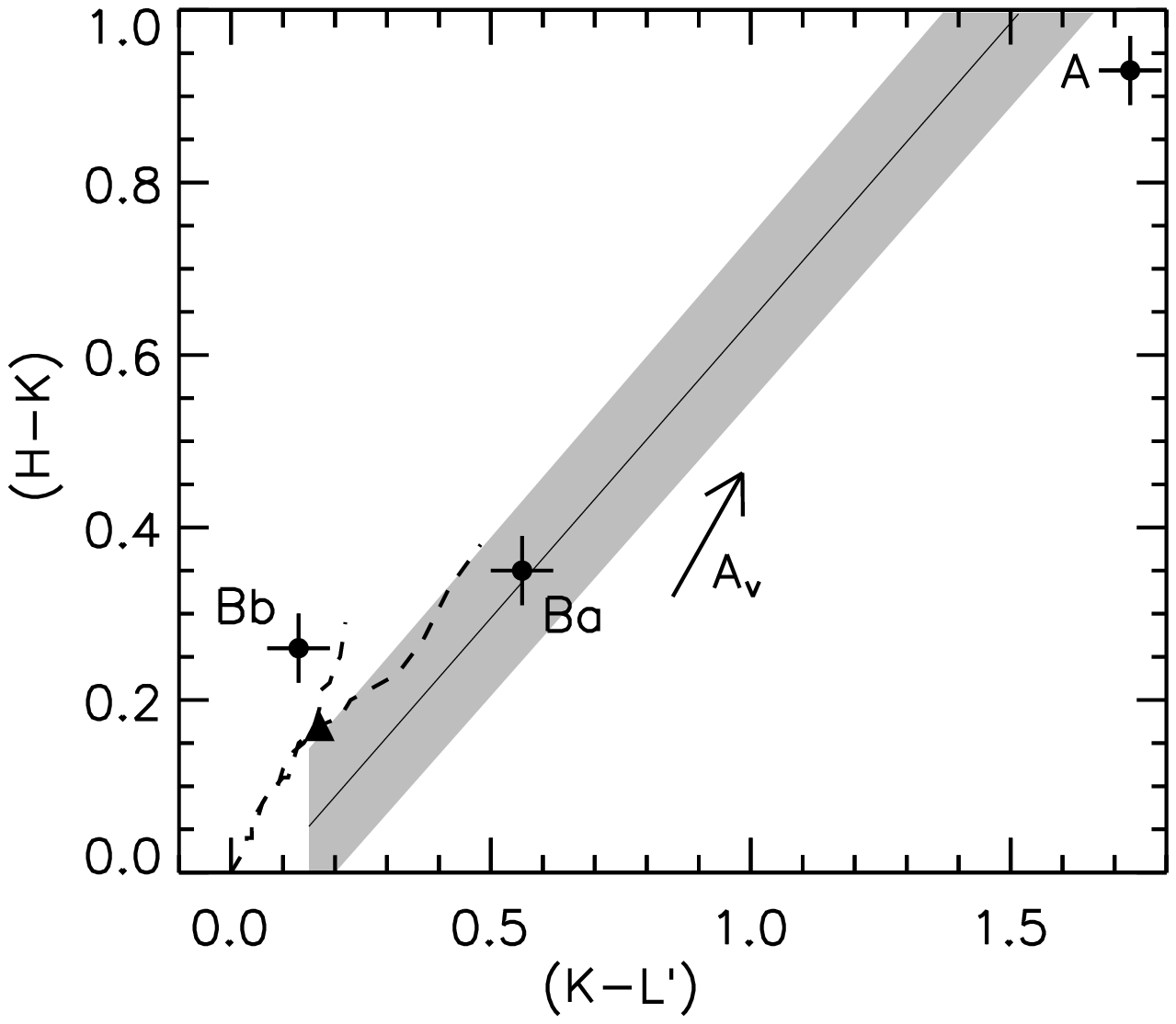}
\caption{AS~353A, Ba, and Bb placed in a color-color diagram showing the 
main-sequence colors of 
dwarfs and giants (dashed lines) and the dereddened classical T Tauri 
star locus (solid line) from Meyer et al. (1997).  
The triangle symbol shows the location of an M0 main-sequence star.
The shaded area shows the approximate width of the scatter about the 
classical T Tauri star locus.
The arrow shows the reddening vector for A$_{\rm{V}}$ = 2.1 in the direction
of increasing extinction.  }
\end{figure}

\vspace{0.5cm}


\begin{thebibliography}{}

\bibitem[]{} Ali, B., Carr, J. S., DePoy, D. L., Frogel, J. A., \& Sellgren,
K. 1995, AJ, 110, 2415

\bibitem[]{} Anglada, G., Villuendas, E., Estalella, R., Beltr\'an,
M. T., Rodr\'\i guez, L. F., Torrelles, J. M. \& Curiel, S. 1998, AJ,
116, 2953

\bibitem[]{} Anosova, J. P. 1986, Ap\&SS, 124, 217

\bibitem[]{} Basri, G., \& Batalha, C. 1990, ApJ, 363, 654.

\bibitem[]{} Bate, M. R., \& Bonnell, I. A. 1997, MNRAS, 285, 33

\bibitem[]{} Bate, M. R., Bonnell, I. A., \& Bromm, V. 2002, MNRAS, 336, 705

\bibitem[]{} B\"ohm, K.-H., \& Raga, A. C. 1987, PASP, 99, 265

\bibitem[]{} Bonnell, I. A., Bate, M. R., Clarke, C. J., \& Pringle, J. E. 2001,
MNRAS, 323, 785

\bibitem[]{} Chelli, A., Cruz-Gonzalez, I., \& Reipurth, B. 1995, 
A\&AS, 114, 135

\bibitem[]{} Cohen, M., \& Kuhi, L. V. 1979, ApJS, 41, 743

\bibitem[]{} Cohen, M., \& Schwartz, R. D. 1983, ApJ, 265, 877

\bibitem[]{} Curiel, S., Raga, A. C., Raymond, J., Noriega-Crespo,
A., \& Cant\'o, J. 1997, AJ, 114, 2736

\bibitem[]{} Dame, T.M., \& Thaddeus, P. 1985, ApJ, 297, 751

\bibitem[]{} Davis, C. J., Eisl\"offel, J., \& Smith, M. D. 1996, ApJ, 463, 246

\bibitem[]{} Duch\^{e}ne, G., Monin, J.-L., Bouvier, J., \& M\'{e}nard, F. 
1999, A\&A, 351, 954

\bibitem[]{} Edwards, S., \& Snell, R. L. 1982, ApJ, 261, 151

\bibitem[]{} Eisl\"offel, J., Solf, J., \& B\"ohm, K.-H. 1990, A\&A, 237, 369

\bibitem[]{} Fernandez, M., \& Eiroa, C. 1996, A\&A, 310, 143

\bibitem[]{} G\'omez, M., Kenyon, S.J., \& Whitney, B. A. 1997, AJ, 114, 265

\bibitem[]{} Gaessler, W., et al. 2002, Proc. SPIE, 4494, 30

\bibitem[]{} Hartigan, P., Mundt, R., \& Stocke, J. 1986, AJ, 91, 1357

\bibitem[]{} Herbig, G. H. 1974, Draft Catalog of Herbig-Haro Objects,
Lick Obs. Bull. No. 658

\bibitem[]{} Herbig, G. H., \& Jones, B. F. 1983, AJ, 88, 1040

\bibitem[]{} Iriarte, B., \&  Chavira, E. 1956, Bol. Obs. Tonantzintla Tacubaya, 14, 31

\bibitem[]{} Jensen, E. L. N., Mathieu, R. D., \& Fuller, G. A.  1996, ApJ, 458, 312

\bibitem[]{} Kobayashi, N., et al. 2000, Proc. SPIE, 4008, 1056

\bibitem[]{} Krisciunas, K., Sinton, W., Tholen, D., Tokunaga, A., Golisch, W.,
Griep, D., Kaminski, C., Impey, C., \& Christian, C. 1987, PASP, 99, 887

\bibitem[]{} Leggett, S. K., et al. 2002, ApJ, 564, 452

\bibitem[]{} Luhman, K. L. 1999, ApJ, 525, 466

\bibitem[]{} Merrill, P. W., \& Burwell, C. G. 1950, ApJ, 112, 72

\bibitem {}  Meyer, M. R., Calvet, N., \& Hillenbrand, L. A. 1997, AJ, 
114, 288

\bibitem {}  Meyer, M. R., Edwards, S., Hinkle, K. H., \& Strom, S. E. 1998, 
 ApJ, 508, 397

\bibitem[]{} Mundt, R., Stocke, J., \& Stockman, H. S. 1983, ApJ, 265, L71

\bibitem[]{} Prato, L., Greene, T. P., \& Simon, M.  2003, ApJ, 548, 853.

\bibitem[]{} Prato, L., Simon, M., Mazeh, T., Zucker, S., \& McLean,
I. S. 2002, ApJ, 579, L99

\bibitem[]{} Reipurth, B. 2000, AJ, 120, 3177

\bibitem[]{} Reipurth, B., Bally, J., \& Devine, D. 1997, AJ, 114, 2708


\bibitem[]{} Simon, M., Chen, W. P., Howell, R. R., Benson, J. A., \&
Slowik, D. 1992, ApJ, 384, 212


\bibitem[]{} Simons, D., \& Tokunaga, A. T. 2002, PASP, 114, 169

\bibitem[]{} Solf, J., B\"ohm, K.-H., \& Raga, A. C. 1986, ApJ, 305, 795

\bibitem[]{} Sterzik, M. F., \& Durisen, R. H. 1995, A\&A, 304, L9

\bibitem[]{} Takami, H., et al. 1998, Proc. SPIE, 3353, 500

\bibitem[]{} Tokunaga, A. T., et al. 1998, Proc. SPIE, 3354, 512

\bibitem[]{} Tokunaga, A. T., Simons, D., \& Vacca, W. D. 2002, PASP, 
114, 180

\bibitem {} Vacca, W. D., Cushing, M. C., \& Rayner, J. T. 2003, PASP, 
115, 389.

\bibitem {} Wallace, L., \& Hinkle, K. 1997, ApJS, 111, 445 

\bibitem[]{} White, R.J., Hillenbrand, L., Metcher, S., \& Patience, J.
2002, BAAS, 34, 1134

\end{thebibliography}
\end{document}